\def\lesssim{\mathrel{\hbox{\rlap{\hbox{\lower4pt\hbox{$\sim$}}}\hbox{$<$}}}}  
  
\def\gtrsim{\mathrel{\hbox{\rlap{\hbox{\lower4pt\hbox{$\sim$}}}\hbox{$>$}}}}  
  
\def\msun{$M_{\odot}$}  
  
\def\teff{$T_{\rm eff}$~}  
  
\def\lteff{log ${T_{\rm eff}}$}  
  
\def\ll_lsun{log$({L/\rm L_{\odot}})$~}  
  
\def\masa_msun{$M/ \rm M_{\odot}$~}  
  
\def\m_mstar{$M/M_{*}$~}

\def\pp{\mbox{\object{PG 0122+200}}}  
\def\pg{\mbox{PG 1159}}  
\def\rxj{\mbox{\object{RX J2117.1+3412}}}
\def\ppg{\mbox{\object{PG 2131+066}}}
\def\ppt{\mbox{\object{PG 1707+427}}} 
\def\ngc{\mbox{\object{NGC 1501}}} 
\def\pgs{\mbox{\object{PG 1159$-$035}}}

\documentclass{aa} 
\usepackage{graphicx}  
          
\begin{document}  
  
\title{On the systematics of asteroseismological mass determinations of 
PG1159 stars}  
  
\author{L. G. Althaus$^{1,2}$\thanks{Member of the Carrera del Investigador  
Cient\'{\i}fico y Tecnol\'ogico, CONICET, Argentina.},  
A. H. C\'orsico$^{1,2\star}$,  S. O. Kepler$^3$, 
\and M. M. Miller 
Bertolami$^{1,2}$\thanks{Fellow of CONICET, Argentina}}   
  
\offprints{L. G. Althaus}  
  
\institute{  
$^1$   Facultad   de   Ciencias  Astron\'omicas   y   Geof\'{\i}sicas,  
Universidad  Nacional de  La Plata,  Paseo del  Bosque S/N,  (1900) La  
Plata, Argentina.\\   
$^2$ Instituto de Astrof\'{\i}sica La Plata, IALP,  
CONICET-UNLP\\    
$^3$ Instituto de F\'isica, Universidade Federal do Rio 
Grande do Sul, 91501-970 Porto Alegre, RS, Brazil \\ 
\email{acorsico,althaus,mmiller@fcaglp.unlp.edu.ar; kepler@if.ufrgs.br} }  
    
\date{Received; accepted}  
  
\abstract{}{We analyze systematics in the asteroseismological
mass determination methods in pulsating \pg\ stars.}{We compare the
seismic masses resulting from the comparison of the observed mean period 
spacings  
with  the  usually adopted  asymptotic period spacings, 
$\Delta \Pi_{\ell}^{\rm  a}$, and  the average  of the computed period 
spacings, $\overline{\Delta \Pi_{\ell}}$. Computations are based 
on full PG1159  evolutionary models with  stellar masses  
ranging from $0.530$ to  $0.741 M_{\odot}$ that take into account   
the complete evolution of progenitor stars.}
{We conclude that asteroseismology  is
a precise and powerful technique that determines the
masses to a high internal accuracy, but it depends 
on the adopted mass determination method. In particular, 
we find that in the case of pulsating \pg\ stars characterized by
short pulsation periods, like  \ppg\ and  \pp, the employment
of the asymptotic period  spacings overestimates the stellar mass by
about 0.06 \msun\ as compared with inferences from the average of the
period spacings. In this case, the  discrepancy between asteroseismological 
and
spectroscopical masses is markedly reduced when use is made of 
the mean period spacing
$\overline{\Delta \Pi_{\ell}}$ instead of the asymptotic period spacing
$\Delta \Pi_{\ell}^{\rm  a}$.}{}  
    
  
\authorrunning{L. G. Althaus et al.}  
  
\titlerunning{On the systematics of asteroseismological mass determinations of 
PG1159 stars}  
  
\maketitle  
  
   
\section{Introduction}  
\label{intro}  
  
Pulsating PG1159 stars (or GW Virginis) are evolved hot stars that can
pose constraints  to the  stellar evolution theory  of post-Asymptotic
Giant Branch (AGB).  These variable  stars belong to the population of
hydrogen--deficient  objects characterized by  surface layers  rich in
helium, carbon and oxygen (Werner  \& Herwig 2006) that are considered
the  evolutionary  link  between   post-AGB  stars  and  most  of  the
hydrogen-deficient  white dwarfs.  The  origin of  most \pg\  stars is
traced  back  to  the   occurrence  of  post--AGB  thermal  pulses:  a
born-again episode induced either by  a very late thermal pulse (VLTP)
experienced  by  a hot  hydrogen-rich  white  dwarf  during its  early
cooling phase ---  see Herwig et al. (1999),  Bl\"ocker (2001), Lawlor
\& MacDonald (2003),  Althaus et al.  (2005), Miller  Bertolami et al.
(2006), or a late thermal pulse (LTP) during which hydrogen deficiency
is a result of a dredge--up episode.  (see Bl\"ocker 2001). During the
VLTP,  the convection  zone driven  by  the helium  flash reaches  the
hydrogen--rich envelope of the star,  with the result that most of the
hydrogen content is burnt.

About  a   third  of spectroscopic  \pg\  stars   exhibit  
multiperiodic  luminosity
variations  with periods in  the range  $300-3000$ s,  attributable to global
nonradial $g$-modes pulsation (e.g. Quirion et al. 2007). The presence of a
pulsational  pattern in  many \pg\  stars has  allowed  researchers to
infer structural parameters ---  particularly the stellar mass --- and
the evolutionary status of  individual pulsators  --- e.g.  Kawaler \& Bradley
(1994), Kawaler  et al.  (1995),  O'Brien et al.  (1998),  Vauclair et
al.  (2002)  and more recently  C\'orsico \& Althaus  (2006).  Stellar
masses of PG1159 stars can  independently be assessed by comparing the
values of  log $ g $  and \lteff, as inferred  from detailed non--LTE
model  atmospheres (Werner  et  al.  1991),  with  tracks coming  from
stellar evolution  modeling, i. e.  the  spectroscopic mass (Dreizler 
\& Heber 1998,  Werner \& Herwig 2006).  These two different approaches
enable us to compare the derived stellar masses.

Recently, considerable  observational and theoretical  effort has been
devoted  to the  study  of some  pulsating  \pg\ stars.   Particularly
noteworthy is the  work of Fu et al. (2007) who  have detected a total
of 23  frequencies in \pp\ and Costa  et al.  (2007)  who have
enlarged to 198 the total  number of pulsation modes in \pgs,
making  it the  star with  the largest  number  of modes
detected   besides   the  Sun.    Parallel   to  these   observational
breakthroughs, substantial progress in the theoretical modeling of
\pg\ stars has been possible (Herwig et al. 1999, Althaus et al. 2005, 
Lawlor \& Mac Donald 2006). In  this sense, the new generation of \pg\
evolutionary models recently developed  by Miller Bertolami \& Althaus
(2006) (hereinafter  MA06)  has  proved  to be  valuable  at  deriving
structural  parameters of pulsating  \pg\ on  the basis  of individual
period fits --- see C\'orsico et al. 2007a and C\'orsico et al. 2007b,
respectively, for an application  to the hot pulsating \rxj\
and the coolest  member of the class, \pp.  These evolutionary
models  are   derived  from  the  complete   evolutionary  history  of
progenitor  stars  with  different  stellar masses  and  an  elaborate
treatment  of the  mixing and  extramixing processes  during  the core
helium burning and born again  phases.  The success of these models at
explaining both the spread in surface chemical composition observed in
\pg\ stars  and the location  of the GW  Vir instability strip  in the
$\log  T_{\rm eff}-  \log g$  plane  (C\'orsico et  al.  2006) renders
reliability to the inferences drawn from individual pulsating \pg.


As shown in  MA06 the employment of detailed  \pg\ evolutionary models
yields  spectroscopical masses  that are  systematically lower  --- by
about 0.05 \msun\ --- than those derived from hydrogen--rich post--AGB
tracks  (Werner  \& Herwig  2006).   Most  importantly, the  resulting
asteroseismological masses (as inferred from the period spacings) are
usually 10\%  higher than the  new spectroscopical masses,  except for
the hot pulsating PG1159 star  \rxj, the spectroscopical mass of which
is more  than 20\% higher than the  asteroseismological one (C\'orsico
et al.   2007a).  The  mass discrepancy is  a clear indication  of the
uncertainties weighting upon the mass determination methods,
even though the spectroscopic uncertainties are of that order.

In  an attempt to  understand the  persisting discrepancy  between the
asteroseismological  and spectroscopical  masses, Miller  Bertolami \&
Althaus (2007) have recently shown  that previous evolution is not the
dominant factor in shaping hydrogen--deficient post--VLTP tracks. They
conclude that the MA06 \pg\ tracks are robust enough as to be used for
spectroscopical  mass   determinations  of  \pg--type   stars,  unless
opacities in  the intershell  region are strongly  subestimated. Their
results   make   clear  that   the   systematic  discrepancy   between
asteroseismological   and  spectroscopical   masses   should  not   be
attributed to uncertainties in post--AGB tracks; rather, they call for
the   need   of  an   analysis   of   possible   systematics  in   the
asteroseismological mass determination methods.  This is precisely the
core feature of the present work. Specifically, we will concentrate on
the  usually adopted  asymptotic period  spacing approach  (Kawaler et
al.  1995; O'Brien  et al.   1998; Vauclair  et al.   2002; Fu  et al.
2007) used in  most mass  determinations of individual  pulsating \pg\
stars. The advantage  of this approach lies in the  fact that the mean
period spacing  of PG1159 pulsators  depends primarily on  the stellar
mass (Kawaler  \& Bradley 1994; C\'orsico \&  Althaus 2006).  However,
the derivation  of the stellar  mass using the  asymptotic predictions
may  not be  entirely reliable  because  they are  strictly valid  for
chemically homogeneous stellar models, while PG1159 stars are expected
to be  chemically stratified with  strong chemical gradients  built up
during  the progenitor  star life.   We will  show that  this approach
overestimates the seismic  mass for those pulsating \pg\  stars on the
white dwarf  cooling track.   We will also  show that  the discrepancy
between  asteroseismological  and  spectroscopic  masses  is  markedly
alleviated if the average of  the computed period spacings, instead of
the asymptotic  ones, is used. In  the next Section,  we summarize the
seismological tools to  infer the stellar mass from  the observed mean
period spacings. We also describe the evolutionary sequences employed.
In Sect. 3 and \ref{period-spacing} we present our results and compare
them with  other mass  determinations methods. We  close the  paper in
Sect. \ref{conclusions} by summarizing our findings.
  
\section{Numerical tools} 

The  most widely  used approach  to  infer the  seismological mass  of
pulsating  \pg\  stars  lies  in  the asymptotic  predictions  of  the
non--radial pulsation theory (with the notable exception of Kawaler \&
Bradley 1994 and C\'orsico et  al. 2007ab). In the asymptotic limit of
very  high radial  order $k$  ($k \gg  1$, i.  e., long  periods), the
$g$--mode periods of a  chemically homogeneous stellar model for a
given degree  $\ell$ and consecutive  $k$ are separated by  a constant
period spacing  $\Delta\Pi_{\ell}^{\rm a}$  given by (Tassoul  et al.
1990)

\begin{equation}
\Delta\Pi_{\ell}^{\rm a}= \frac{\Pi_0}{\sqrt{\ell(\ell+1)}}= 
\frac{2 \pi^2}{\sqrt{\ell(\ell+1)}} \left[ \int_{r_1}^{r_2}  
(N/r) dr \right]^{-1},
\label{asymp}
\end{equation}

\noindent being $N$ the Brunt-V\"ais\"al\"a frequency given by

\begin{equation}
N^2= -g\left[ \frac{d\ {\rm ln}\ \rho}{dr} - \frac{1}{\Gamma_1}
\frac{d\ {\rm ln}\ P}{dr} \right],
\label{brunt}
\end{equation}

\noindent where $g$ is the local gravity and $\Gamma_1$ the
first adiabatic exponent (see Hansen \& Kawaler 1994).
Note that the term in brackets is the difference between the
real and the adiabatic density gradients, that determines buoyancy.  The integral is
taken   over   the    $g$--mode   propagation   region.    Note   that
$\Delta\Pi_{\ell}^{\rm a}$ is a  function of the structural properties
of the  star via the Brunt-V\"ais\"al\"a  frequency. The seismological
stellar    mass   is    constrained    by    directly    comparing
$\Delta\Pi_{\ell}^{\rm a}$ as computed  from Eq.  \ref{asymp} with the
observed mean period spacing if the effective temperature
of the  target star is known (by means  of spectroscopy).  Full
advantage is taken of the fact that the $g$--mode period spacing of
\pg\ pulsators is mostly sensitive to the stellar mass and only weakly
dependent  on the  stellar luminosity  and helium--rich  envelope mass
fraction  (Kawaler \& Bradley  1994). This  feature together  with the
fact that no detailed  pulsation calculations are required   to   compute
$\Delta\Pi_{\ell}^{\rm a}$ turns the  asymptotic period spacing into a
 practical tool  to infer the  stellar mass  of pulsating
\pg\ stars.

\begin{figure}  
\centering  
\includegraphics[clip,width=250pt]{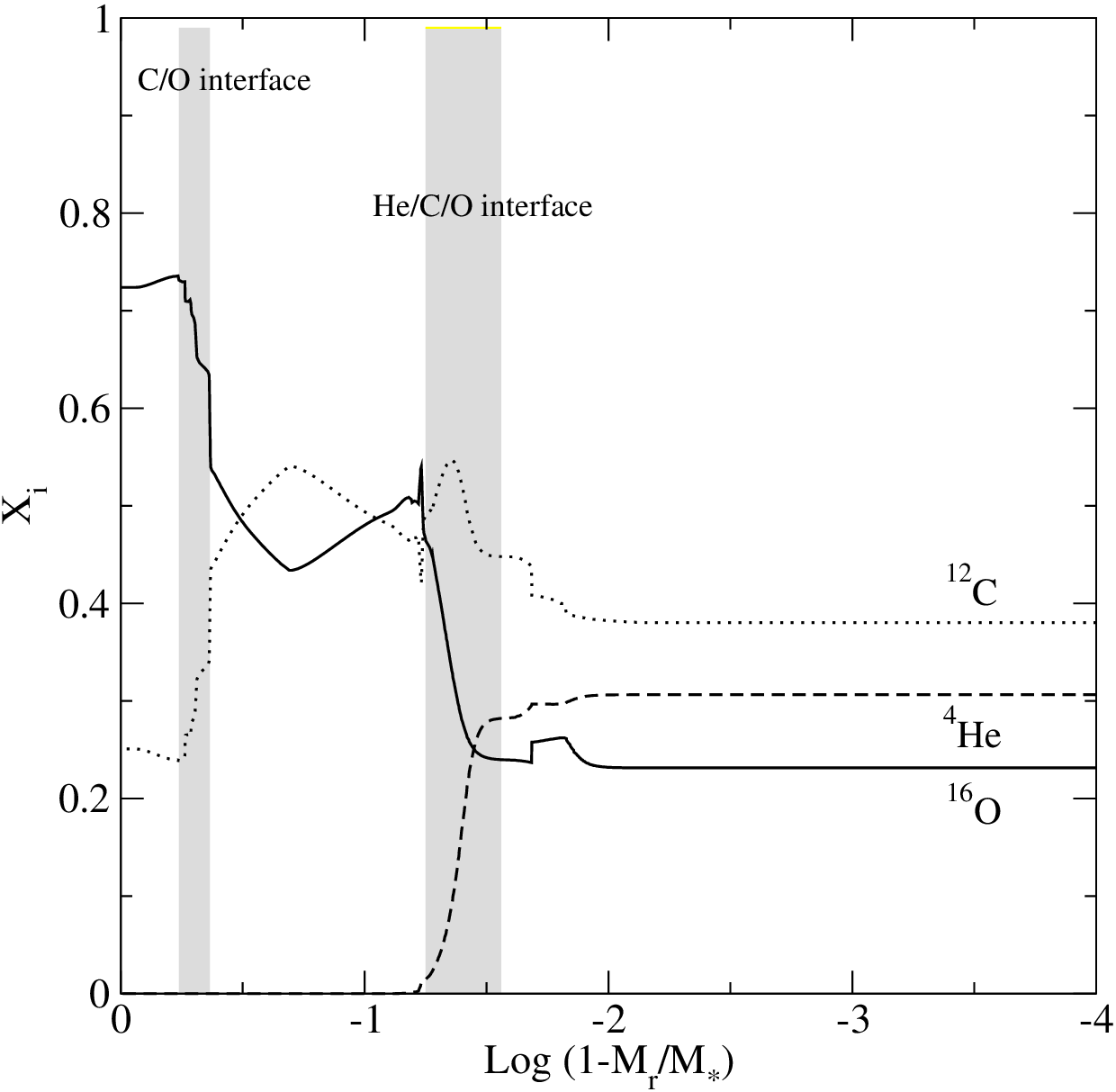}  
\caption{The inner chemical abundance distribution corresponding
to a 0.589 \msun\ \pg\ model at \lteff= 5.18. The approximate 
locations of the C/O and He/C/O chemical transition regions are 
emphasized with gray.}
\label{quimi}  
\end{figure}  

As mentioned, the  asymptotic  formula  given   by  Eq.  \ref{asymp} 
is  strictly valid  for  chemically  homogeneous
stellar models and in the limit of high $k$, i.e. long periods.  
However, according to the current
theory of stellar evolution, \pg\  stars are expected to be chemically
stratified characterized by  strong  chemical transitions built
up  during   the  progenitor  star   life.  This  is   illustrated  by
Fig.   \ref{quimi}  which  displays   the  inner   chemical  abundance
distribution in  a typical \pg\  star. Two main  chemical transitions,
emphasized  with gray, are  easily recognized:  an  inner C/O
interface  left  by the  extra  mixing  episodes  that ocurred  during
central  helium burning  (see Straniero  et al.   2003) and  an He/C/O
interface   that   separates  the   helium--rich   envelope from   the
carbon--oxygen core ---  modeled by nuclear processing in prior AGB
and  post--AGB stages.   Such  chemical interfaces  produce clear  and
distinctive signatures in $N$, which are critical for the mode--trapping
properties  of  the  models.  These mode  trapping  features  strongly
disturb  the  structure  of  the  period spectrum,  thus  causing  the
computed $g$--mode  period spacing ($\Pi_{k+1} - \Pi_{k}$)
to appreciably depart from uniformity (see Kawaler \& Bradley 1994 and
more recently C\'orsico \& Althaus 2005, 2006).

A more realistic approach to infer the stellar mass of \pg\ stars that
does not suffer  from the  above mentioned shortcomings  is to  compare the
observed  period  spacing with  the  average  of  the computed  period
spacings, $\overline{\Delta  \Pi_{\ell}}$.  This quantity  is assessed by
averaging the  computed forward  period spacings in  the same range  as the
observed periods, that is

\begin{equation}
\overline{\Delta \Pi_{\ell}}= \frac{1}{n} 
\sum_k \Delta  \Pi_{k}=  \frac{1}{n} \sum_k
(\Pi_{k+1} - \Pi_{k})
\label{average}
\end{equation}

\noindent where $n$ means the number of observed modes (with $m$=0) of 
the star.  In  contrast with  the
asymptotic  approach, the assessment  of the  asteroseismological mass
via  $\overline{\Delta \Pi_{\ell}}$  involves the  computation  of the
full  adiabatic  period spectrum.  Accurate  values  of the  adiabatic
pulsation periods  of pulsating \pg\ stars requires  the employment of
{\sl full
\pg\ evolutionary models} that  reflect the thermal structure of their
progenitors  (Kawaler  et al.  1985).  In  this  work, we  employ  the
evolutionary models  recently developed by  Althaus et al.  (2005),
MA06, and C\'orsico et al. (2006, 2007b) who computed the complete evolution
of model  star sequences with initial  masses on the  ZAMS (assuming a
metallicity  of $Z= 0.02$) in the  range 1  --- 3.75~$M_{\odot}$. These
authors  have  followed all  of  the sequences
through the thermally pulsing and  mass--loss phases on the AGB to the
\pg\ regime.   The evolutionary  stages corresponding to  the complete
burning of protons shortly after the occurrence of the VLTP and the ensuing  
born-again episode that give rise to the
H-deficient, He-,  C- and O-rich composition  characteristic of PG1159
stars have  been carefully followed  for each sequence. The  masses of
the resulting  remnants span the range 0.530 ---
$0.741~M_{\odot}$. For  these \pg\ evolutionary sequences  
we have computed $\ell= 1$, $g$-mode adiabatic pulsation periods
with the same  numerical code and methods employed in those
works (see C\'orsico \& Althaus  2006 for details).  
In what follows,  we will use these evolutionary  models to compute 
both the mean  $\overline{\Delta \Pi_{\ell}}$ and
asymptotic $\Delta\Pi_{\ell}^{\rm a}$.

\section{Discrepancy between  the asymptotic and
the average of period spacings}  

\begin{figure}  
\centering  
\includegraphics[clip,width=244pt]{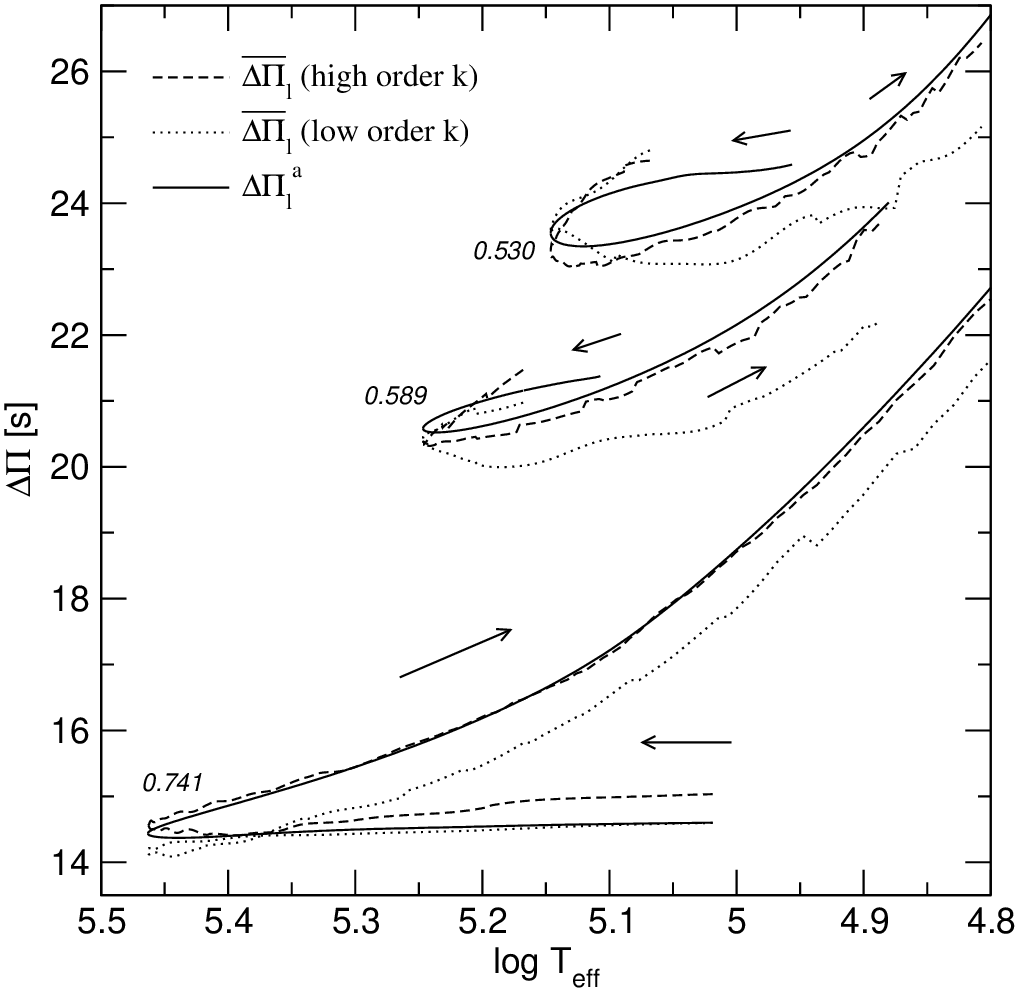}  
\caption{The dipole ($\ell =1$) asymptotic period spacing 
($\Delta\Pi_{\ell}^{\rm a}$, 
solid line) is compared with the average of period spacing, 
$\overline{\Delta \Pi_{\ell}}$, as a function of the effective temperature 
for the 0.53, 0.589 and 0.741 \msun\ evolutionary sequences. For
 $\overline{\Delta \Pi_{\ell}}$ we consider short and long periods, 
i.e., low and high $k$ values (dotted and dashed lines, respectively). 
Stages  before and after the models 
reach their highest effective temperature are shown.  Arrows indicate
the direction of evolution.}
\label{compare}  
\end{figure}

Here we employ the  evolutionary models described previously to assess
the asymptotic  period spacing,  $\Delta \Pi_{\ell}^{\rm a}$,  and the
average of the computed period spacings, $\overline{\Delta
\Pi_{\ell}}$ as given by Eqs. \ref{asymp} and \ref{average}, respectively. 
In Fig. \ref{compare},  which summarizes the main result  of our work,
we show the  run of these two quantities for $\ell=  1$ modes
(as most detected periodicities are triplets) in terms
of the  effective temperature for selected stellar  masses.  To assess
the dependence  of $\overline{\Delta \Pi_{\ell}}$ on  the period range
where  the  average  of  the   period  spacing  is  done,  we  compute
$\overline{\Delta \Pi_{\ell}}$ for intervals of short and long periods
(300--600 s   and   900--1500 s,  respectively). Different stars have 
different ranges.    The resulting
$\overline{\Delta \Pi_{\ell}}$ in each  case are denoted by dotted and
dashed  lines.   Note  that   both  $\Delta  \Pi_{\ell}^{\rm  a}$  and
$\overline{\Delta \Pi_{\ell}}$ decrease as the stellar mass increases.
Note also  that, when the star  evolves along the  white dwarf cooling
track,  the   period  spacings  increase   with  decreasing  effective
temperature.  This is due to  the increasing degeneracy in the core as
the star cools, causing the Brunt-V\"ais\"al\"a frequency to gradually
decrease, and the consequent slow increment in the periods.

Most  importantly,  note  from  Fig.  \ref{compare}  that,  generally,
$\overline{\Delta \Pi_{\ell}}$ turns out to be smaller than $\Delta
\Pi_{\ell}^{\rm   a}$.    Note   also   the   marked   dependence   of
$\overline{\Delta  \Pi_{\ell}}$ on  the  period interval  where it  is
calculated.   Indeed, $\overline{\Delta  \Pi_{\ell}}$ may  be markedly
distinct from the $\Delta \Pi_{\ell}^{\rm a}$ predictions depending on
the range  of periods in  which the average  of the period  spacing is
performed (or observed).   This is  particularly  true for  the 
evolutionary  stages
corresponding to  the white dwarf  regime, where, for a  given stellar
mass, $\overline{\Delta \Pi_{\ell}}$ turns out to be about 1 s smaller
than  $\Delta \Pi_{\ell}^{\rm  a}$ when  averages are  taken  on short
period  intervals.  It  is  apparent that  only  in the  case of  long
periods do  the period spacings  given by $\Delta  \Pi_{\ell}^{\rm a}$
resemble those predicted  by $\overline{\Delta \Pi_{\ell}}$, i.e.  the
asymptotic conditions  are nearly  reached in this  case.  In  view of
this, we expect that for those pulsating \pg\ stars on the white dwarf
cooling  track,  that usually  exhibit  short  pulsation periods,  the
stellar  mass  inferred  from $\overline{\Delta  \Pi_{\ell}}$  becomes
substantially smaller  than the  stellar mass determined  from $\Delta
\Pi_{\ell}^{\rm  a}$.    We  address  this  issue   in  the  following
section. On  the other  hand, for the  stages before  the evolutionary
knee, the mean $\overline{\Delta \Pi_{\ell}}$ values tend to be larger than
the asymptotic $\Delta \Pi_{\ell}^{\rm a}$ ones.

\begin{figure} 
\centering 
\includegraphics[clip,width=250pt]{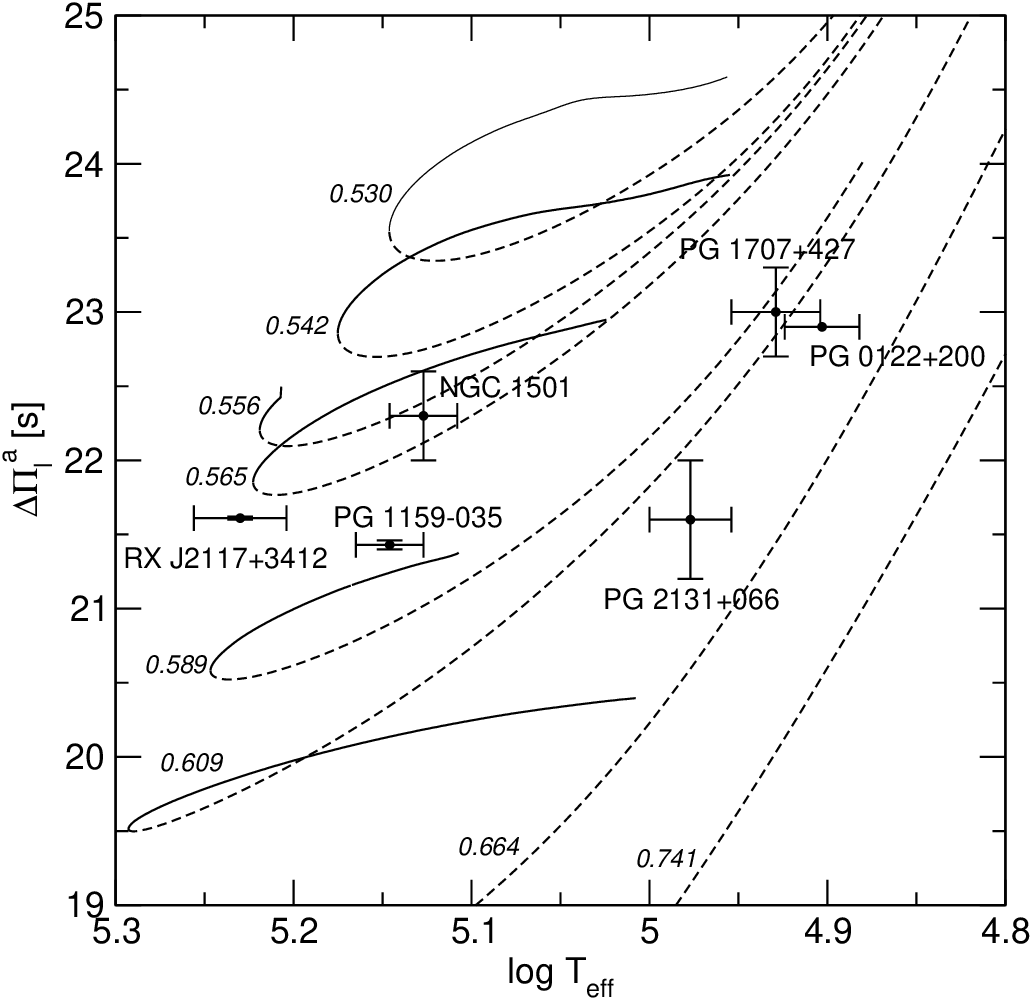} 
\caption{The dipole ($\ell =1$) asymptotic period spacing 
($\Delta  \Pi_{\ell}^{\rm   a}$)  in terms  of  the  effective
temperature  for  various   stellar  masses.   Solid  (dashed)  lines
correspond  to stages before  (after) the  models reach  their highest
effective temperature (evolutionary knee). Also, the location of pulsating
\pg\ stars with observed mean period spacings is shown. See Table 1
for details.}
\label{asint}  
\end{figure}

\section{Mass determinations from the observed period spacings}  
\label{period-spacing}  

Here we  employ the evolutionary models described  previously to infer
the seismic  mass of  selected pulsating \pg\  stars by  comparing the
asymptotic  period  spacing,  $\Delta  \Pi_{\ell}^{\rm  a}$,  and  the
average of the computed period spacings, $\overline{\Delta
\Pi_{\ell}}$, with the {\it observed} mean period spacing, $\Delta \Pi^{\rm
O}$. These methods  allow us to
infer a value of the stellar mass as long as the effective temperature
of the star is determined from spectroscopy or other method. Naturally 
one parameter,
$\Delta \Pi$, cannot determine two properties, $\mathrm{Teff}$ and $\log g$.

\begin{figure}  
\centering  
\includegraphics[clip,width=250pt]{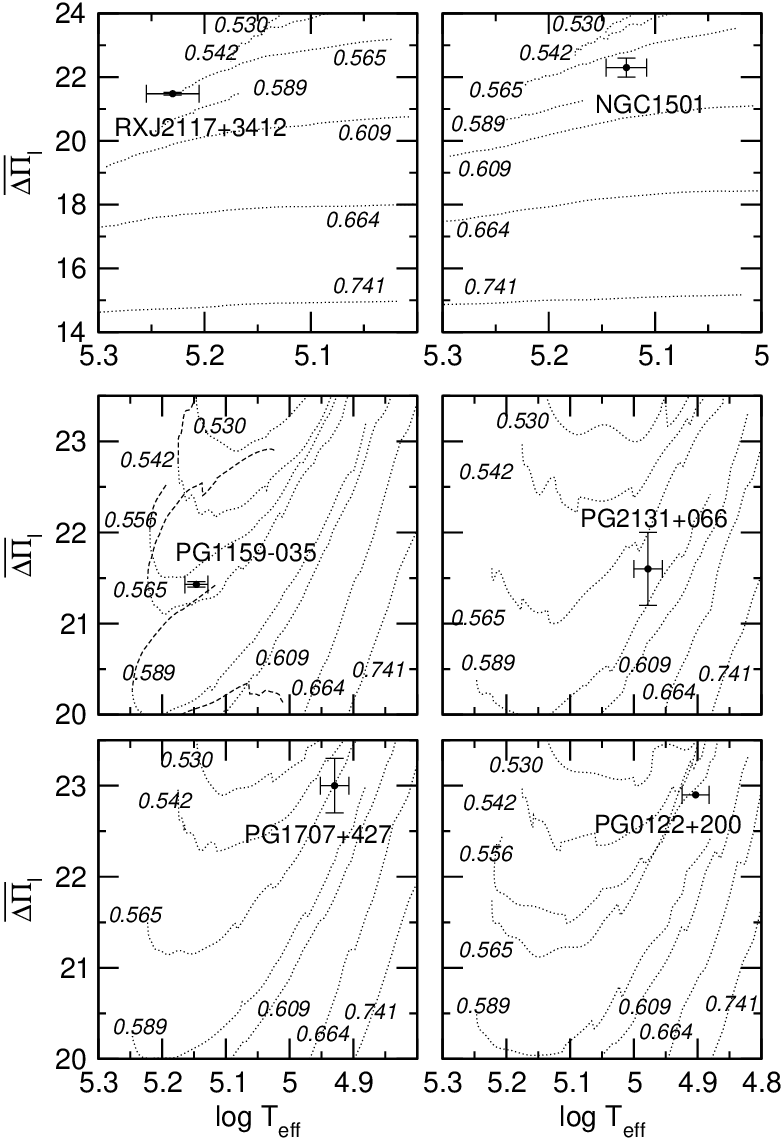}  
\caption{The average of the computed period spacings for \pg\ model
sequences with different stellar masses in terms of the effective temperature. 
Each panel corresponds to a specific pulsating \pg. Also, the observed
mean period spacings are shown. The top two panels correspond to 
evolutionary stages before the sequences reach the maximum effective 
temperature,
i.e., in the PNN stage instead of the DO stage.}
\label{panels}  
\end{figure} 

In Fig. \ref{asint}, we  show the evolution of $\Delta \Pi_{\ell}^{\rm
a}$ (for $\ell =1$) in terms of the effective temperature for the MA06
\pg\  evolutionary  models.   The  predictions  corresponding  to  the
evolutionary  stages  before  the  maximum effective  temperature  are
indicated with  solid lines,  while the stages  tracing the
later evolution,  hot white
dwarf cooling branch, are denoted  with dashed lines. In addition, the
location of  pulsating \pg\ stars with the  most recent determinations
of  the  observed  mean  period  spacings, $\Delta  \Pi^{\rm  O}$,  are
included in Fig.  \ref{asint} --- and also listed  in the sixth column
of Table 1. Specifically, we include the observational data for
\ppg, \pp, \ppt, \rxj, \pgs,
and \ngc. These pulsating stars are hot hydrogen--deficient, post--AGB
stars for which the number  of pulsation modes detected is high enough
to infer an average of the period spacings.  \ppg,
\pp, and  \ppt\ are evolved \pg\ stars  on the hot white
dwarf  cooling  branch  and  characterized by  {\sl  short}  pulsation
periods (see last column in Table 1). By contrast, \rxj\ and
\ngc,  low--gravity and  high--luminosity  objects, pulsate  with
markedly   longer  periods.   The pulsating \ngc\ belongs to the
[WCE] class, Wolf Rayet--type  central stars of planetary nebulae with
emission lines and believed to be the progenitors of \pg\ stars.

From the asymptotic $\Delta \Pi_{\ell}^{\rm a}$ diagram shown in 
Fig. \ref{asint},
the  stellar mass  of  the  above mentioned  pulsating  \pg\ stars  is
assessed.  The results are  listed in  the second  column of  Table 1.
Note  that  the  seismic  masses  as  inferred from  the  use  of  the
asymptotic approach  differ by more than 10\% 
from  the spectroscopic masses
(the spectroscopic masses are taken  from MA06 and listed in the fifth
column  of Table  1)\footnote{We mention that in the case
of \pgs\ and \pp\, the uncertainty in the measured surface gravity translates
into an uncertainty of $\pm 0.1$ \msun\  in the spectroscopic mass (MA06).}. 
This  difference  is particularly  true for  the
short--period  variables such  as  \pp,  \ppg, and \ppt, for  which the seismic  mass becomes about 18\%  larger than
the spectroscopic one.

\begin{table*}[] 
\centering
\caption{Stellar masses for selected pulsating \pg\ stars as derived
from the asymptotic and the average period spacings (second and third 
columns). The fourth column lists the stellar mass resulting from
detailed period fittings, when available. The fifth columns displays 
the stellar mass as inferred from spectroscopy (from  \teff and $g$ values 
from Werner \& Herwig 2006). The sixth columns corresponds to the observed 
period spacings and the last column the observed range of
periods for $\ell=1$.  All masses are in solar units.}
\begin{tabular}{ccccccc}
\hline
Star          & $M_*[\Delta  \Pi_{\ell}^{\rm  a}]$ & $M_*[\overline{\Delta  
\Pi_{\ell}}]$ & $M_*[\rm fit]$ & $M_*[\rm spectr]$ 
&  $\Delta \Pi^{\rm
O} $ & Obs. period range \\
 & This work &  & & MA06 & [s] & [s]\\ \hline
PG 2131+066	   &	0.627 &	0.578  &  & 0.55 & 21.6$^d$ & 339-598$^i$	\\   
PG 0122+200	   &	0.625 & 0.567$^a$  & 0.556$^a$ & 0.53 & 22.90$^e$ & 335-611$^e$ \\  
PG 1707+427        &	0.597 &	0.566  &  & 0.53 & 23.0$^f$ & 335-909$^f$	\\
RX J2117.1+3412    &	0.568 &	0.560$^b$ & 0.565$^b$ & 0.72 & 21.62$^g$ & 694-1530$^g$ \\
PG 1159$-$035	   &	0.577-0.585$^{**}$ &	0.561$^c$ & 0.565$^c$ & 0.54 & 21.43$^h$ & 390-990$^h$ \\   
NGC 1501	   &	0.571 &	0.576 &  & 0.56 & 22.3$^j$ & 1154-2000$^j$ \\ 
\hline    
\end{tabular} 
\label{tab:masitas} 

{\footnotesize   References: (a)  C\'orsico et al. (2007b);  (b)  
C\'orsico et al. (2007a); (c) C\'orsico et al. (2007c); 
(d) Reed et al. (2000); (e) Fu et al. (2007); (f) Kawaler et al. (2004);
(g) Vauclair et al. (2002); (h) Costa et al. (2007); (i) Kawaler et al. (1995);
(j) Bond et al. (1996). (**) The two mass values result from considering 
that the
star is either after or before the evolutionary knee.}
\end{table*}

From the discussion in the  previous section, we expect for our target
stars  smaller stellar  masses when  they  are derived  from the  mean
$\overline{\Delta \Pi_{\ell}}$. This is borne out by Fig \ref{panels},
which  displays $\overline{\Delta \Pi_{\ell}}$  for $\ell=1$  modes in
terms of the effective temperature for different stellar masses.  Each
panel corresponds to a specific  star, discussed above.  To derive the
average  of  the  period  spacings  in \rxj\  and  \ngc,  we  computed
$\overline{\Delta \Pi_{\ell}}$  for the high--luminosity  (PNN) regime
of the evolutionary sequence models,  while for the remaining stars we
compute  values  of  $\overline{\Delta  \Pi_{\ell}}$  for  the  stages
following  the  evolutionary  knee   for  the  \pg\  stars,  i.e.  the
low--luminosity   (DO)  regime.   Also   for  each   star,  the   mean
$\overline{\Delta  \Pi_{\ell}}$ is calculated  by averaging  the model
period spacings  over the corresponding  period interval in  which the
periodicities are  indeed observed.  This  is the reason for  the fact
that the  curves are different in  each panel. In the  third column of
Table 1 we  list the resulting estimation of the  stellar mass for the
six stars.  For those pulsating \pg\ characterized  by short pulsation
periods,  the   seismic  masses  as  derived  by   this  approach  are
appreciably  lower --- up  to 0.06  \msun\ lower  --- than  the values
inferred  by using the  asymptotic period  spacing.  As  we mentioned,
this is due to the mean $\overline{\Delta \Pi_{\ell}}$ being typically
0.7$-$1.0 s  smaller than  the asymptotic $\Delta  \Pi_{\ell}^{\rm a}$
when  short   periods  are  involved,   i.e.  for  stages   after  the
evolutionary  knee.    Thus,  the  discrepancy   between  seismic  and
spectroscopic masses  is markedly alleviated  when the average  of the
period  spacings is  used  instead the  asymptotic  ones. Indeed,  the
seismic  mass in  this  case becomes  at  most 6  \%  larger than  the
spectroscopically derived  masses, except  for the hot  pulsating \pg\
star \rxj, the spectroscopical mass  of which is more than 20\% higher
than the asteroseismological mass.

\section{Discussion and conclusions}  
\label{conclusions}  

This paper explores the systematic discrepancy between spectroscopical
and asteroseismological masses of pulsating
\pg\ stars.  Our  motivation is the result  of Miller Bertolami
\& Althaus  (2007) that such  discrepancy should not be  attributed to
uncertainties in post--AGB tracks,  but possibly to systematics in the
asteroseismological mass determination  methods. Recently, Quirion has
pointed to  one of us (M3B)  that a possible  opacity change resulting
from the spread of He/C/O abundances in PG1159 stars could be a source
of uncertainty in the location of the tracks.  We addressed this issue
by calculating sequences in which helium and carbon are changed in the whole
envelope above the helium burning  shell. We find that changing helium
into carbon by an amount of 0.4  by mass shifts the track by only 0.02
dex in effective  temperature (being bluer if carbon  is higher). This
translates  into  a shift  of  only 0.005  and  0.015  \msun\ for  the
spectroscopic mass near the  0.51 and 0.6 \msun\ tracks, respectively.
Thus, the  precise values  of the He/C/O  abundances do not  seem to
introduce appreciable changes in the masses derived by MA06.

Specifically, we  have concentrated on the seismic  masses that result
from a  comparison of  the observed period  spacings with  the usually
adopted asymptotic period  spacings ($\Delta \Pi_{\ell}^{\rm a}$) used
in  most mass  determination  of individual  pulsating  \pg\, and  the
better    suited   average   of    the   computed    period   spacings
($\overline{\Delta \Pi_{\ell}})$.
On  the basis  of  full  \pg\ evolutionary  models  that consider  the
evolutionary  history  of  progenitor  stars  (MA06),and  the  ensuing
internal chemical  profile, we have  shown that the derivation  of the
stellar mass using the asymptotic period spacing is not appropriate in
the case of \pg\ stars.   In particular, we demonstrate that for those
pulsating \pg\ stars characterized  by short pulsation periods, i. e.,
the pulsating  \pg\ stars  on the hot  white dwarf regime  (DOVs), the
asymptotic  $\Delta \Pi_{\ell}^{\rm a}$  differs appreciably  (by more
than 1 s)  from the mean $\overline{\Delta \Pi_{\ell}}$.   Only in the
case   of   variables   with    long   periods   (PNNVs),   like   the
high--luminosity, log--gravity pulsating \pg\  stars, do the $g-$ mode
period  spacings  given  by  asymptotic  $\Delta  \Pi_{\ell}^{\rm  a}$
resemble those predicted by mean $\overline{\Delta \Pi_{\ell}}$.  This
is expected  because the asymptotic  conditions are approached  in the
limit of very high radial order $k$.

For  quantitative  inferences,  we  have  computed  the  seismic  mass
resulting from the employment of the asymptotic and the average of the
computed  period  spacing  for  those  pulsating  \pg\  which  have  a
sufficiently large number of detected modes to infer an observed value
of the mean period spacing. Our  selected stars are listed in Table 1,
together  with the  stellar  mass inferences.    
The employment of the asymptotic theory, in principle formally valid for 
chemically homogeneous stellar models at high radial index k, 
overestimates the seismic mass by about 0.06 \msun\ in the case
of  very short period  pulsating \pg\  stars like \ppg\  and \pp.   
Because \pg\ stars  are  expected to  be chemically  stratified,
estimations  of the stellar  mass from mean  $\overline{\Delta \Pi_{\ell}}$
are  more  realistic than  those  inferred by  means of 
asymptotic $\Delta \Pi_{\ell}^{\rm    a}$.    Indeed,    stellar   mass    
derived   from the mean
$\overline{\Delta  \Pi_{\ell}}$ are  in good  agreement with  the mass
values obtained  from detailed  period fittings. The  discrepancy
between  asteroseismological   and  spectroscopical  masses   is  markedly
alleviated by  the employment  of the average  of the  computed period
spacing instead of the asymptotic period spacings.

In  closing,  a  Fortran  program  to derive,  from  our  evolutionary
sequences,  averages  of  the  period  spacing  for  arbitrary  period
intervals      is       available      at      our       web      site
http://www.fcaglp.unlp.edu.ar/evolgroup.

\begin{acknowledgements}  
We acknowledge an anonymous referee for the comments about our
paper.
This research was partially supported by IALP and  PIP 6521 grant   
from CONICET.  
\end{acknowledgements}

\end{document}